\documentclass[english,aps,twocolumn]{revtex4-1}
\usepackage[T1]{fontenc}
\usepackage[latin9]{inputenc}
\usepackage{color}
\usepackage{amsbsy}
\usepackage{amstext}
\usepackage{graphicx}

\makeatletter

\@ifundefined{textcolor}{}
{%
 \definecolor{BLACK}{gray}{0}
 \definecolor{WHITE}{gray}{1}
 \definecolor{RED}{rgb}{1,0,0}
 \definecolor{GREEN}{rgb}{0,1,0}
 \definecolor{BLUE}{rgb}{0,0,1}
 \definecolor{CYAN}{cmyk}{1,0,0,0}
 \definecolor{MAGENTA}{cmyk}{0,1,0,0}
 \definecolor{YELLOW}{cmyk}{0,0,1,0}
}

\makeatother

\usepackage{babel}
\begin{document}
\global\long\def\ket#1{\left|#1\right\rangle }

\title{Snowflake Topological Insulator for Sound Waves}

\author{Christian Brendel}

\affiliation{Max Planck Institute for the Science of Light, Staudtstra{ß}e 2,
91058 Erlangen, Germany}

\author{Vittorio Peano}

\affiliation{Department of Physics, University of Malta, Msida MSD 2080, Malta}

\author{Oskar Painter}

\affiliation{Institute for Quantum Information and Matter and Thomas J. Watson,
Sr., Laboratory of Applied Physics, California Institute of Technology,
Pasadena, USA }

\author{Florian Marquardt}

\affiliation{Max Planck Institute for the Science of Light, Staudtstra{ß}e 2,
91058 Erlangen, Germany}

\affiliation{Institute for Theoretical Physics, University of Erlangen-Nürnberg,
Staudtstr. 7, 91058 Erlangen, Germany}
\begin{abstract}
We show how the snowflake phononic crystal structure, which has been
realized experimentally recently, can be turned into a topological
insulator for sound waves. This idea, based purely on simple geometrical
modifications, could be readily implemented on the nanoscale.
\end{abstract}
\maketitle
\emph{Introduction}. \textendash{} First examples of topologically
protected sound wave transport have just emerged during the past three
years. So far, experimental implementations exist on the centimeter-scale,
both for the case of time-reversal symmetry broken by external driving
\cite{nash_topological_2015}, such as in coupled gyroscopes, as well
as for the case without driving \cite{susstrunk_observation_2015,he_acoustic_2016,lu_observation_2016,ye_observation_2016},
such as in coupled pendula. Moreover, a multitude of different implementations
have been envisioned theoretically \cite{prodan_topological_2009,wang_topological_2015,yang_topological_2015,kariyado_manipulation_2015,khanikaev_topologically_2015,chen_tunable_2016,fleury_floquet_2016,pal_helical_2016,lu_valley_2016,matlack_designing_2016,mei_pseudo-time-reversal_2016,kariyado_mirror_2016,pal_edge_2016,susstrunk_classification_2016}.
However, it is highly desirable to come up with alternative design
ideas that may be realized on the nanoscale, eventually pushing towards
applications in integrated phononics. The first theoretical proposal
in this direction \cite{peano_topological_2015} suggested to exploit
the optomechanical interaction to imprint the optical vorticity of
a suitably shaped laser beam to generate chiral sound wave transport
in a phononic-photonic crystal. On the other hand, if one wants to
avoid the strong driving by an external field, purely geometrical
designs are called for. One remarkable idea of Mousavi et al. \cite{mousavi_topologically_2015}
posited creating a sound wave topological insulator by designing a
phononic crystal structure made from a material that would be carefully
engineered by a pattern of small holes to achieve degeneracy between
vibrations that are symmetric and antisymmetric to the plane of the
sample. The appearance of a fine-grained length-scale much smaller
than the wavelength, however, makes it impossible to use this idea
all the way down to wavelengths comparable to the smallest feature
sizes allowed by nanofabrication. In the present manuscript, we propose
a very simple modification to an already existing structure, the so-called
snowflake phononic crystal. The snowflake crystal has already proven
to be a reliable platform for nanoscale optomechanics \cite{safavi-naeini_two-dimensional_2014},
and could also support pseudomagnetic fields for sound waves \cite{brendel_pseudomagnetic_2016}.
With the proposed modification, which is inspired by an idea first
analyzed by Wu and Hu for photonic systems \cite{wu_scheme_2015}
(see also \cite{barik_two-dimensionally_2016,kariyado_mirror_2016,mei_pseudo-time-reversal_2016}),
one will be able to create a topological insulator for sound waves
based on a proven nanoscale platform. 

The envisaged system consists of snowflake-shaped holes of alternating
sizes, in a periodic arrangement on a triangular lattice. This snowflake
topological insulator can be viewed as a metamaterial that supports
topologically protected sound waves whose typical wavelength is larger
than the underlying lattice scale. Such elastic waves propagate along
arbitrarily shaped domain walls engineered by appropriately varying
the snowflake size. We will show that the topological protection is
guaranteed if locally (at the lattice scale) the point group symmetry
of the snowflake design is mantained. \textcolor{red}{}

\begin{figure}
\includegraphics[width=1\columnwidth]{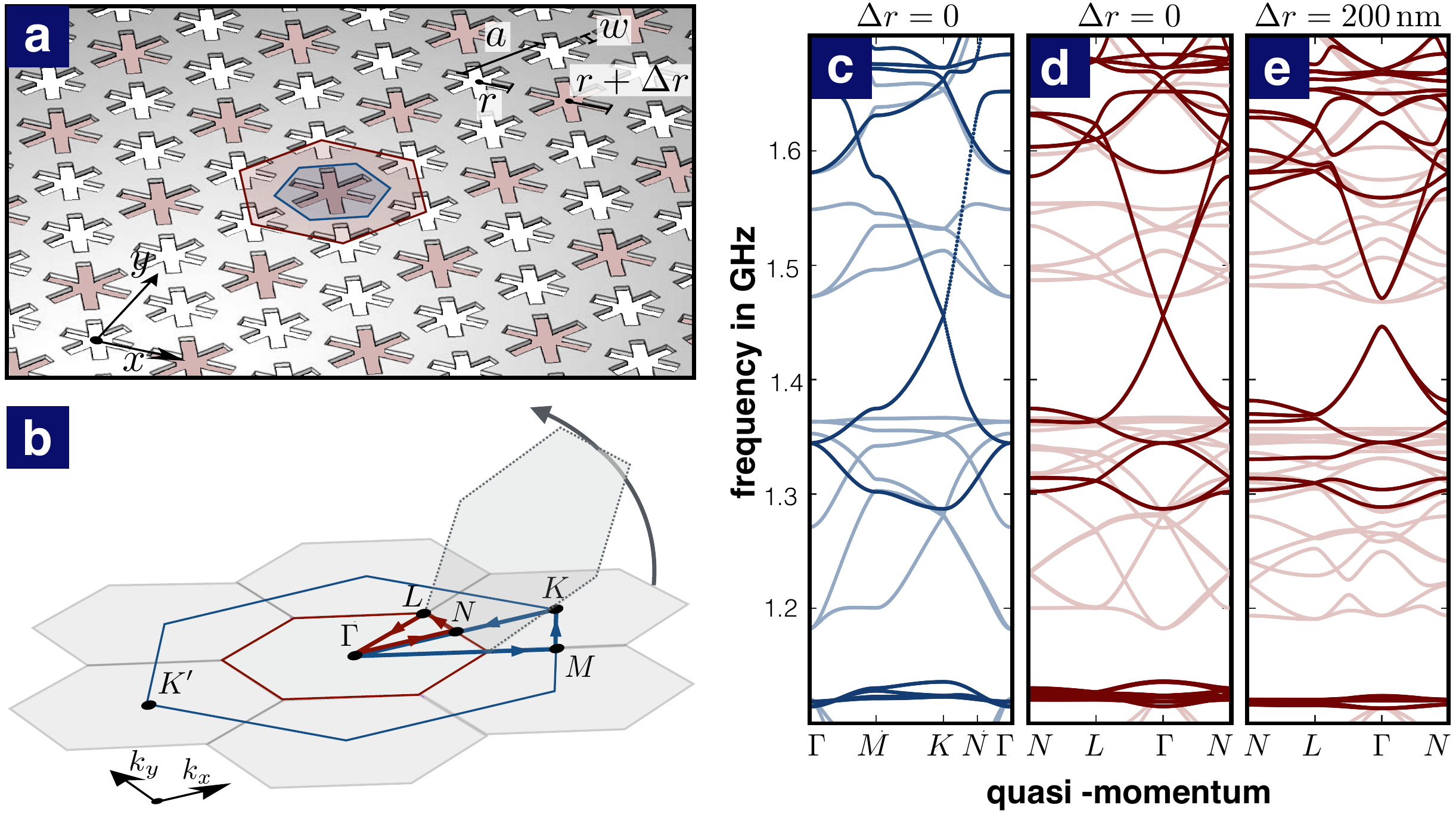}

\protect\caption{\label{Fig1}(a) Snowflake crystal slab defined by the parameters
$(a,r,w,\Delta r)$ and the slab's thickness $d$. $\Delta r$ indicates
the radius deviation of every third (red) snowflake. For the special
case of $\Delta r=0$ we obtain a regular snowflake pattern with the
discrete translational symmetry $T_{a}$ (blue unit cell). (c) shows
the corresponding band structure along a path passing the high symmetry
points (blue BZ in b) that features Dirac cones at $\vec{K}$ and
$\vec{K}'$. Describing the same system with an enlarged unit cell
(red) results in a reduced BZ (red), folds the band structure, and
maps both Dirac valleys to the $\vec{\Gamma}$-point (d). $\Delta r\protect\neq0$
breaks $T_{a}$ but maintains $T_{\sqrt{3}a}$ (requiring the red
unit cell and red BZ) and $\mathcal{C}_{6}$. This gaps both Dirac
double cones (e). In the band structures (c-e) modes symmetric to
the $x\text{-}y\text{-}$plane are displayed in brighter colors. {[}Here
we used $(a,r,w,d)=(5000,1800,750,220)\,\text{nm}$ and assumed the
crystal slab made of silicon (Si) with Young\textquoteright s modulus
of $170\,\text{GPa}$, mass density $2329\,\text{kg/m}^{3}$ and Poisson\textquoteright s
ratio $0.28${]}}
\end{figure}

\emph{Platform}.\textbf{ }\textendash{} We assume a planar quasi-two-dimensional
phononic crystal slab exhibiting a six-fold rotational symmetry ($\mathcal{C}_{6}$)
as well as a discrete translational symmetry ($T_{a}$) on a triangular
lattice, with a lattice constant $a$. The most straightforward implementation
consists in the snowflake phononic crystal. This crystal has been
explored before in the context of optomechanics \cite{safavi-naeini_two-dimensional_2014},
since it is both a photonic and a phononic crystal, although we will
only make use of its phononic properties. Its structure is shown in
Fig.~\ref{Fig1}a.

\emph{Symmetries and folding}. \textendash{} Due to the $\mathcal{C}_{6}$-symmetry,
the acoustic band structure is forced to have Dirac cones at the two
high-symmetry points, $\vec{K}=\pi/a(4/3,0)$ and $\vec{K}'=\pi/a(2/3,2/\sqrt{3})$.
Now consider a single snowflake-shaped hole surrounded by six other
such holes. Our aim will be to break the original translational symmetry
by changing the central snowflake in this configuration, thereby enlarging
the real-space unit cell by a factor of $\sqrt{3}$ (Fig.~\ref{Fig1}a).
Conversely, this will reduce the size of the first Brillouin zone
(BZ) by the same factor. To anticipate this reduction, we imagine
what happens when the original band structure, obtained for the as-yet
unperturbed structure, gets folded back into the new BZ (see Fig.~\ref{Fig1}a-b).
This will map the Dirac cones from $\vec{K}$ and $\vec{K}'$ of the
old BZ to the $\Gamma$-point of the new BZ, forming a degenerate
pair of double Dirac cones at $\vec{\Gamma}=(0,0)$ (Fig.~\ref{Fig1}c-d).
Only in the next step we actually do break $T_{a}$, by carrying out
the afore-mentioned change of the central snowflake, increasing the
lattice constant $a\rightarrow\tilde{a}=\sqrt{3}a$ and establishing
the new symmetry \textbf{$T_{\tilde{a}}$}, all the while preserving
$\mathcal{C}_{6}$ around the unit-cell centers. This opens a complete
gap at the $\Gamma$-point (Fig.~\ref{Fig1}e), which can be topological
in nature, as will be explained below.

The easiest way to implement this modification of the geometry is
to change the radius of the central snowflake in each (enlarged) unit
cell by $\Delta r$, with $\Delta r=0$ referring to the original
snowflake crystal (Fig.~\ref{Fig1}a). 

The band structures shown in Fig.~\ref{Fig1} have been obtained
from full finite-element simulations of the equations of linear elasticity,
solving the eigenvalue problem 
\begin{equation}
\text{div}\Big[{\bf E}:\left[\text{grad}{\bf \boldsymbol{\psi}}+(\text{grad}{\bf \boldsymbol{\psi}})^{T}\right]\Big]=-2\varrho\omega^{2}\boldsymbol{\psi},
\end{equation}
with $\boldsymbol{\psi}$ being the complex three-dimensional wave
function related to the mechanical displacement field ${\bf u}=\text{Re}\left[\boldsymbol{\psi}\cdot e^{i\omega t}\right]$
of the crystal, ${\bf E}$ the elasticity tensor, and $\varrho$ the
mass density. Here $:$ is a short-hand for the tensor product, $[\mathbf{E}:{\rm grad}\psi]_{ij}=E_{ijkl}\partial_{l}\psi_{k}$.
For concreteness, we envisioned a phononic crystal slab of thickness
$220\,{\rm nm}$, but much smaller slabs have been fabricated already,
down to unit cells of only a few hundred nanometers lateral extension
(acoustic frequencies would scale inversely with the linear dimensions,
as usual).

\emph{Effective Hamiltonian.} \textendash{}\textbf{ }We now derive
an effective Hamiltonian valid for the vicinity of the $\Gamma$-point.
We are taking a route that clarifies the connection to the original
valley degree of freedom. The results can alternatively be understood
in the framework of the symmetry arguments first advocated for $\mathcal{C}_{6}$-symmetric
structures in the photonic context in \cite{wu_scheme_2015}. We start
with the eigenstates of the two double Dirac cones of the regular
($\Delta r=0$) snowflake array. They are labeled by $\ket{\psi_{\sigma,\tau}}$,
with $\sigma=\pm$1 being the quasi-angular momentum with respect
to the 3-fold rotation
\begin{equation}
\hat{R}_{2\pi/3}\ket{\psi_{\sigma,\tau}}=e^{-i\sigma\frac{{2\pi}}{3}}\ket{\psi_{\sigma,\tau}},
\end{equation}
and $\tau$ denoting the valley degree of freedom ($\tau=+1$ for
$K$ vs. $\tau=-1$ for $K'$ ). Under time reversal $\hat{T}$ (complex
conjugation) and inversion $\hat{R}_{\pi}$ they obey 
\begin{equation}
\ket{\psi_{\sigma,\tau}}=\hat{T}\ket{\psi_{-\sigma,-\tau}}=\hat{R}_{\pi}\ket{\psi_{\sigma,-\tau}}.
\end{equation}
The exact mode shapes of $\ket{\psi_{\sigma,\tau}}$ for the particular
case of the original snowflake crystal are explained in detail in
\cite{brendel_pseudomagnetic_2016}. In order to derive the Hamiltonian,
we use two sets of Pauli matrices to span the 4-dimensional Hilbert
space. We define one set for the valley degree of freedom $\hat{\tau}_{\{x,y,z\}}$
and another one for the quasi angular degree of freedom $\hat{\sigma}_{\{x,y,z\}}$,
such that $\hat{\tau}_{z}\ket{\psi_{\sigma,\tau}}=\tau\ket{\psi_{\sigma,\tau}}$
and $\hat{\sigma}_{z}\ket{\psi_{\sigma,\tau}}=\sigma\ket{\psi_{\sigma,\tau}}$,
and the usual set of Pauli matrices holds in this basis.

We now write down the Hamiltonian as a Taylor series up to linear
order in $\vec{k}$ by using the above matrices. We keep only terms
that are invariant under $\hat{T}$ and $\hat{R}_{\pi/3}$, which
are the symmetries of our snowflake crystal (even for $\Delta r\neq0$).
This leaves us with only the following terms: 
\begin{equation}
\hat{H}_{{\bf k}}=g\hat{\tau}_{x}+v\hat{\tau}_{z}(k_{x}\hat{\sigma}_{x}+k_{y}\hat{\sigma}_{y}).\label{eq:Hk}
\end{equation}
Up to a unitary transformation this Hamiltonian is equivalent to the
large-wavelength limit of the Bernevig-Hughes-Zhang model for a topological
insulator \cite{bernevig_quantum_2006}. The first term in Eq. (\ref{eq:Hk})
is induced by the breaking of the $\hat{T}_{a}$ symmetry and is responsible
for gapping the degenerate Dirac cones. In other words, $g$ can be
interpreted as a mass, which can change sign. The unitary symmetry
that allows to cast Eq. (\ref{eq:Hk}) in a block-diagonal form  is
the spin degree of freedom $\hat{S}=\hat{\tau}_{x}\hat{\sigma}_{z}$
. Combined with the time-reversal operator, it gives rise to a pseudo
time-reversal symmetry $(\hat{T}\hat{S})$, which has the peculiarity
that it squares to minus the identity, directly leading to Kramer's
degeneracy. At the $\Gamma$-point, the common eigenstates of $\hat{H}_{{\bf k}=0}$
and $\hat{S}$ are the states $\ket{p^{\pm}}$ and $\ket{d^{\pm}}$
which obey $\hat{\tau}_{x}\ket{d^{\pm}}=\ket{d^{\pm}}$, $\hat{\tau}_{x}\ket{p^{\pm}}=-\ket{p^{\pm}}$, $\hat{S}\ket{p^{\pm}}=\pm\ket{p^{\pm}}$,
and $\hat{S}\ket{d^{\pm}}=\pm\ket{d^{\pm}}$.  One can show that these
states are actually of $p$- and $d$-type with respect to their behavior
under 60-degree rotations: 
\[
\hat{R}_{\pi/3}\ket{p^{\pm}}=e^{\pm i\pi/3}\ket{p^{\pm}},\quad\hat{R}_{\pi/3}\ket{d^{\pm}}=e^{\pm i2\pi/3}\ket{d^{\pm}}\,.
\]
Note that away from the $\Gamma$ point only states of the same helicity
($s=\pm1$) will get mixed to form the finite$-k$ eigenstates.\textcolor{red}{}The
Hamiltonian terms which are linear in the quasimomentum $\vec{k}$
will induce transitions only between states whose $60$-degree quasi-angular
momenta differ by one quantum: $p^{+}$ to $d^{+}$ and $p^{-}$ to
$d^{-}$. Only further away from the $\Gamma$-point, higher-order
terms (e.g. $\sim k^{2}$) can eventually couple the $+$ and $-$
states, i.e. mix different helicities. An indirect signature of this
coupling is the lifting of the degeneracy \textcolor{black}{of the
two helicities.}\textcolor{red}{}\textcolor{black}{{} Remarkably, for
our specific design, the splitting remains smaller than $1.5\%$ of
the band gap even for a quasimomentum as large as 1/4 of the distance
to the boundary of the Brillouin zone, $|\vec{k}|\leq\pi/(6a)$.}

\begin{figure}
\begin{centering}
\includegraphics[width=1\columnwidth]{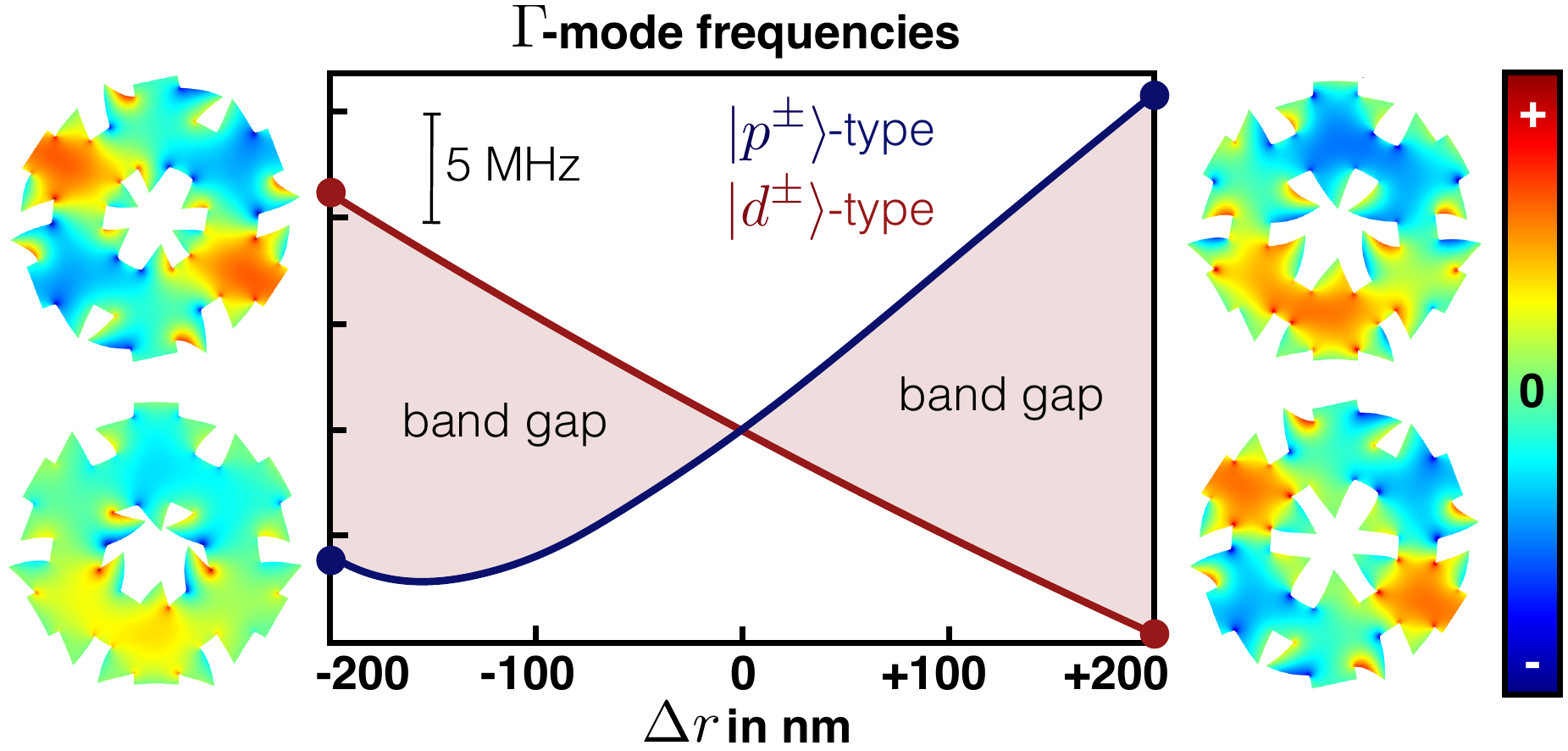}
\par\end{centering}

\protect\caption{\label{Fig2}Band inversion: Frequencies of the $|p^{\pm}\rangle,|d^{\pm}\rangle$
modes at the $\Gamma$-point $\vec{k}=0$ evolving for a sweep of
the central snowflake radius $r_{c}$. A snapshot of the corresponding
displacement fields for $r_{c}=1600$nm and $r_{c}=2000$nm is also
shown. The in-plane displacement field is directly visualized by the
deformation, whereas the out-of-plane displacement is encoded in the
colorscale. $d$($p$)-orbitals are (anti-)symmetric under rotation
by 180 degrees.}
 
\end{figure}

In a topological insulator, the helical edge states are confined along
domain walls that separate regions of opposite mass $g$. Here, we
can simply tune the mass $g$ by changing the radius of the central
snowflake, see Fig. \ref{Fig2}. As discussed above when all snowflakes
have the same radius (corresponding to $\Delta r=0$) the $p-$ and
$d-$ bands are degenerate at the $\Gamma$ point ($g=0$). For a
decreased (increased) radius $r_{c}$ of the central snowflakes, the
$d-$ orbitals have larger (smaller) energy, corresponding to a positive
(negative) mass $g$, cf. Fig. \ref{Fig2}. In order to understand
this behavior, it is useful to observe that the $p$- orbitals have
extra nodes at the external links leading out of the (enlarged) unit
cell, enforced by a phase-difference of $\pi$ across those links.
When all snowflakes have equal radius, the additional energy cost
associated with the larger phase gradient (compared to a $d$-orbital)
across these external links exactly offsets the benefit of a reduced
phase gradient on a path encircling the central snowflake. Obviously,
stronger (weaker) internal links {[}corresponding to a decreased (increased)
central snowflake radius $r_{c}${]} favor energetically the $p$-
($d$-) states, eventually leading to the behavior displayed in Fig.
\ref{Fig2}.

\emph{Strip in the continuum model.} \textendash{}\textbf{ }To ascertain
the appearance of edge states at an interface with a band inversion,
we first consider a strip configuration with a mass term $g(y)$ varying
along the transverse (finite) $y$-direction from $-g_{0}$ to $+g_{0}$.
We employ the continuum limit based on Eq.~(\ref{eq:Hk}), using
the envelope function approximation. Following this standard procedure
\cite{hasan_topological_2010,asboth_short_2016,jackiw_solitons_1976},
we obtain a right (left) moving state, with a linear dispersion $E=vk_{x}$
($E=-vk_{x}$) for $s=-1$ ($s=+1$), that decays exponentially away
from the domain wall, with a penetration depth $\xi=v/g_{0}$. This
behaviour, obtained in the continuum limit, is confirmed in direct
finite-element simulations of the microscopic equations of elasticity
for a strip geometry.

\begin{figure}
\includegraphics[width=1\columnwidth]{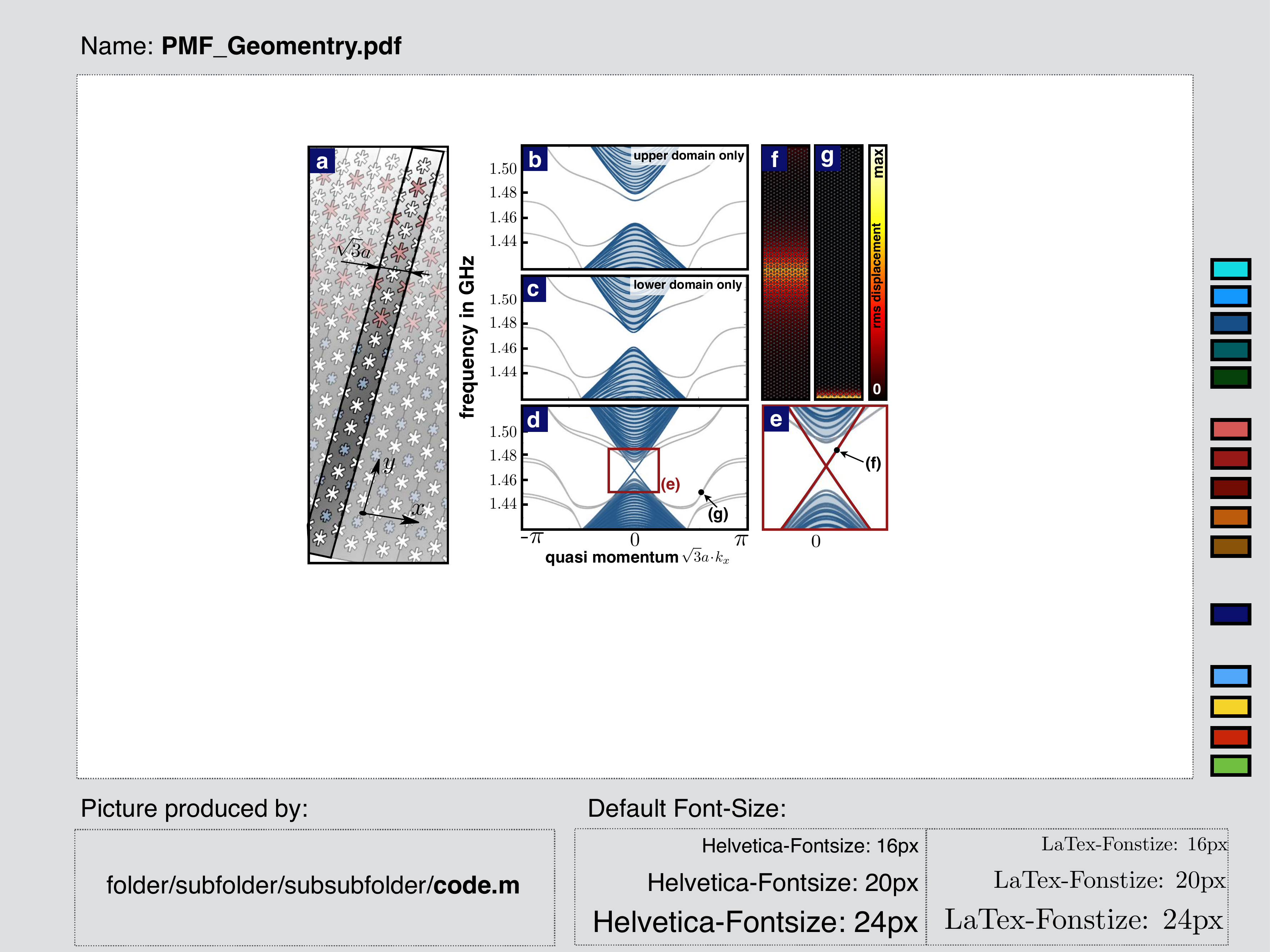}

\protect\caption{\textcolor{blue}{\label{Fig3}}(a) Snowflake strip configuration comprising
two different domains. The lower domain has smaller central snowflakes
(blue, $\Delta r=-200\,\text{nm}$) whereas the corresponding ones
in the upper domain are enlarged (red $\Delta r=+200\,\text{nm}$).
The depicted strip hosts $n=11$ resized snowflakes ($n_{l}=6$ blue
and $n_{u}=5$ red). (b) Acoustic band structure of the blue domain
only, with a system size $n=17$, i.e. a strip comprising just one
domain ($\Delta r=-200\,\text{nm}$). The blue area shows the band
structure of the corresponding infinite crystal, thereby indicating
the strip's bulk bands, whereas the sole remaining bands are degenerate
pairs of edge states localized at the upper~($u$) and lower~($l$)
boundary of the strip. (c) Counterpart to (b) for the upper domain
($\Delta r=+200\,\text{nm}$), essentially the strip version of Fig.~\ref{Fig1}e.
A strip comprising two domains ($\Delta r_{l/u}=\mp200\,\text{nm}$,$n_{l}=18$,$n_{u}=17$)
has the band structure depicted in (d,e). In addition to the bulk
modes and the edge modes at the physical boundaries (g), it reveals
the two topologically protected counterpropagating modes localized
at the domain wall where the gap closes {[}wave function in (f){]}.
For clarity, in all these band structures we just depicted the modes
symmetric to the sample plane.}
\end{figure}

\emph{Helical edge channels in finite-element simulations}.\textbf{
}\textendash{} We will now verify the above statements for the snowflake
crystal, using the full microscopic acoustic equations. For that purpose,
we consider a strip with a finite extent along $y$. Before we investigate
the effects of domain walls, we first briefly discuss the strip with
a spatially homogeneous mass term, i.e. composed of the hexagonal
building blocks comprising three snowflakes (Fig.~\ref{Fig1}a, red
shaded area), with the central snowflake's radius deviating by $\Delta r$.
Figure \ref{Fig3} shows the band structures of strip configurations
with $\Delta r=-200\,\text{nm}$ (b) and $\Delta r=200\,\text{nm}$
(c), obtained by the COMSOL finite element solver. The Dirac cones
are replaced by a complete bulk band gap. There are states that appear
in addition to the bulk-derived bands and that are localized at the
boundaries (Figure \ref{Fig3}e), arising due to the symmetry-breaking
at these sharp sample boundaries. These edge states are not protected
by any symmetry and are highly sensitive to the exact geometry of
the edge. Moreover, they are two-fold degenerate; one state is localized
at the upper and the other at the lower boundary.

Next, we attach both structures to each other (Fig.~\ref{Fig3}a)
and obtain a strip geometry with a domain wall where the sign of the
mass term $g\sim\Delta r$ flips. The corresponding band structure
is shown in figure \ref{Fig3}d, which is basically a superposition
of the band structures of the bare strips with $\Delta r=\pm200\,\text{nm}$.
However, in addition to the bulk bands and the afore-mentioned states
at the sample boundaries, two states appear that traverse the gap
entirely, with a linear dispersion of opposite slope (group velocity).
Moreover, there is no discernible avoided crossing between these two
states, underlining the absence of back-scattering expected for topological
insulators due to the symmetry-protection. Figure \ref{Fig3}e shows
the quasi-momentum resolved wave function of the right-moving state
(red energy dispersion in panel d). For small quasi-momenta it is
highly confined around the domain wall, with a typical penetration
depth inversely proportional to the size of the bulk band gap (as
expected from $\xi=v/g_{0}$ derived in the continuum model).

\begin{figure}
\begin{centering}
\includegraphics[width=1\columnwidth]{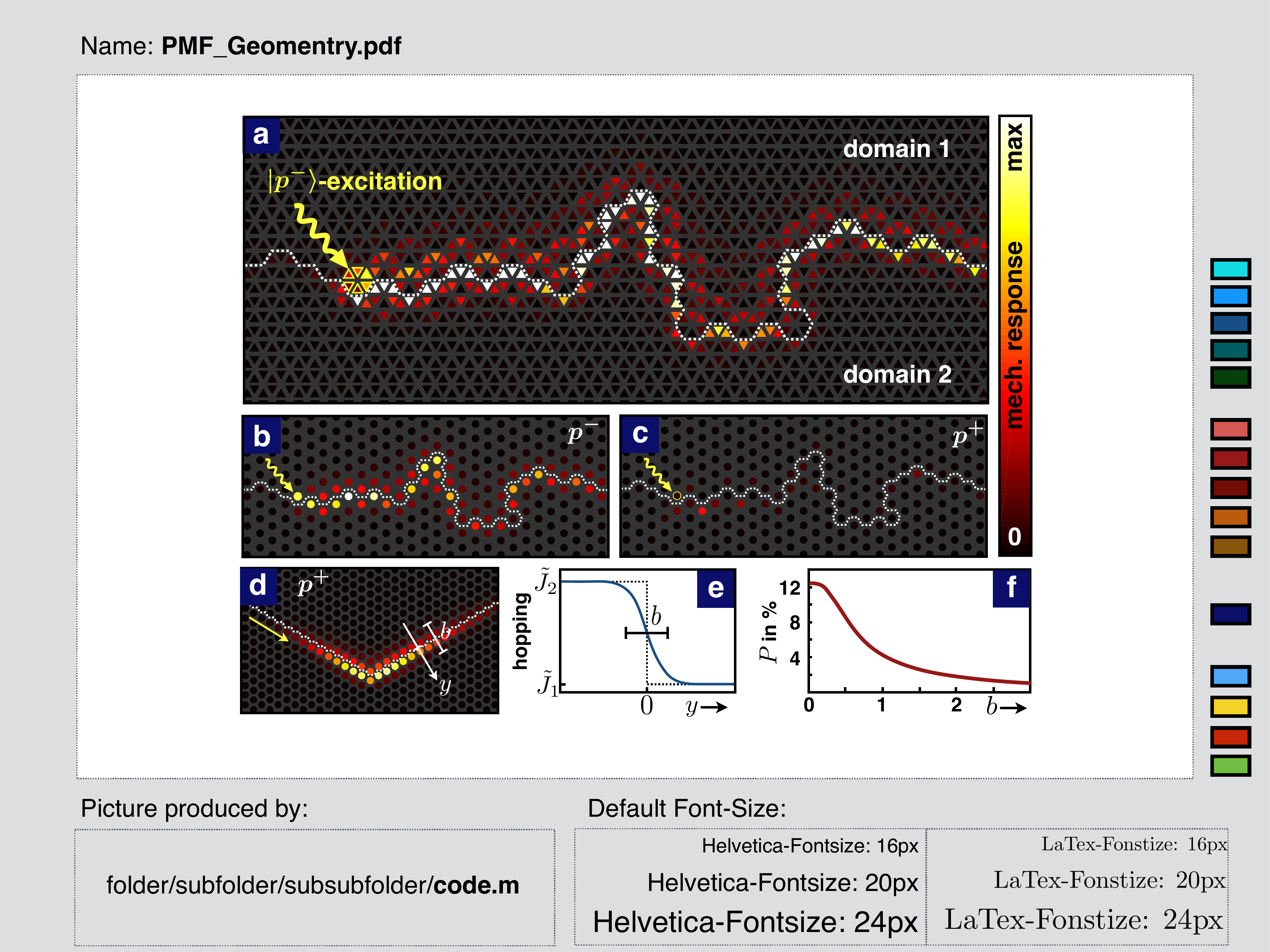}
\par\end{centering}

\protect\caption{\label{Fig4}Finite size sample with arbitrarily shaped sharp domain
wall, simulated using a tight-binding model. One polarization ($p^{-}$)
is injected and propagates to the right. (a) Detailed representation,
showing all 6 triangles inside each hexagonal unit cell. Color indicates
the square of the sound wave amplitude, i.e. the energy. (b) Extracting
the component of the $p^{-}$ mode inside each unit cell. (c) Scattering
into the other helicity ($p^{+}$) is present, but still strongly
suppressed even at sharp corners. Note that in panel (c) we enhanced
the depicted energy by a factor of 10 to make the weak scattered component
visible. (d) Weak component of opposite polarization ($p^{+}$) appearing
at a corner {[}color scale different from before{]}. (e) Domain wall
geometry (transverse to corner). (f) The total fraction of $p^{+}$
polarization decreases for a smoother domain wall.}
\end{figure}

\emph{Effects of disorder}. \textendash{} The engineered symmetry
$\hat{S}$ will not protect against competely arbitrary (generic)
disorder. This behaviour is in fact common to all bosonic topological
insulators \cite{lu_topological_2014}. However, $\hat{S}$ is valid
near the $\Gamma$-point. Therefore, generically speaking, one may
expect that this protection is preserved near smooth defects, interfaces,
and sample boundaries, which admix only wavevectors close to $\Gamma$.
On top of this, even for sharp domain boundaries, we find in the full
finite-element simulations that there is no discernible back-scattering
for the parameters we have explored. Such a scattering would show
up in the form of a minigap, i.e. an avoided crossing between the
counterpropagating edge states. This unexpected robustness in the
presence of sharp interfaces has been observed as well in works analyzing
photonic structures based on $C_{6}$ symmetry \cite{wu_scheme_2015,barik_two-dimensionally_2016}.
Moreover, one can show that even short-range defects of a certain
kind, that are compatible with the symmetry $\hat{S}$, do not induce
scattering between counterpropagating edge states. For example, both
of the following perturbations have zero matrix element between sectors
of opposite helicity: changing the masses of all 6 triangles in a
given unit cell by the same amount, or leaving out a single snowflake
hole.

\emph{Arbitrary boundaries.} \textendash{} We now consider a finite
system with an arbitrarily shaped boundary to observe unidirectional
transport of mechanical excitations and the effects of disorder. To
keep the computational effort manageable, we approximate the snowflake
crystal by a tight-binding model, where the sites of this model directly
correspond to the physical triangles that are arranged in a honeycomb
lattice. Note that in order to obtain most of the insights we are
aiming for here, it would be more generally sufficient to use any
tight-binding model that exhibits the same symmetries that underly
the topological protection in our system. We use a hexagonal unit
cell comprising six sites with equal eigenfrequencies. To mimic the
mass term, hopping rates within a unit cell, $\tilde{J}$, deviate
from hopping rates between different unit cells, $J$. Again there
is a direct interpretation: By changing the radius of the central
snowflake in the real microscopic structure, the links connecting
the triangles also change, consequently leading to different coupling
between two triangles.

As shown above, the unidirectional edge states are superpositions
of the states with the same helicity $s$ (e.g. $p^{+}$ and $d^{+}$).
Figure \ref{Fig4} shows the energy distribution for a mechanical
wave that is propagating at the domain wall between two domains with
different mass terms. We excite a whole unit cell (indicated by the
yellow arrow) with a $|p^{-}\rangle$-type mode shape, thereby launching
a sound wave that just propagates to the right. By calculating the
linear response of each lattice site to this particular excitation,
we obtain the propagation probability (modulus squared of the Green's
function) of the mechanical excitation. As mentioned in the general
discussion near Eq.~(\ref{eq:Hk}), the symmetry $\hat{S}$ is obeyed
to a very good approximation within a rather large fraction of the
Brillouin zone. As a consequence, any remaining admixture of the opposite
helicity is quickly suppressed when a sharp domain wall (with a sudden
jump in hopping amplitudes) is replaced by even only a slightly smoothened
wall. This is confirmed by the numerical simulations displayed in
Fig.~\ref{Fig4}f. 

\emph{Conclusions}. \textendash{} The snowflake topological insulator
for sound waves proposed here is straightforward to fabricate at any
scale, down to the nanoscale. It can be excited and read-out using
a variety of different approaches, including electrical, mechanical,
and optomechanical (adapting the ideas presented in \cite{brendel_pseudomagnetic_2016}).
The simplicity of the nanoscale design (and the small dimensions of
the unit cell) will turn such a modified snowflake crystal into a
versatile platform for generating arbitrary phononic circuits and
networks \cite{habraken_continuous_2012,schmidt_optomechanical_2012}
on the chip, which may couple to hybrid quantum systems of various
kinds and could also contain optically tuneable non-reciprocal elements
\cite{fang_generalized_2017}.

\emph{Acknowledgments}. \textendash{} We acknowledge funding by ERC
StG OPTOMECH and the EU HOT network, as well as the Max Planck Society.
O.P. acknowledges support by the AFOSR-MURI Quantum Photonic Matter,
the ARO-MURI Quantum Opto-Mechanics with Atoms and Nanostructured
Diamond (grant N00014-15- 1-2761), and the Institute for Quantum Information
and Matter, an NSF Physics Frontiers Center (grant PHY-1125565) with
support of the Gordon and Betty Moore Foundation (grant GBMF-2644).\textbf{ }

\bibliographystyle{apsrev4-1}
\bibliography{ChiralSound}

\end{document}